\documentstyle[12pt]{article}
\begin{document}
\title{New physics effects in $\epsilon'/\epsilon$ }
\author{Yue-Liang Wu\footnote{Supported in part by
 Outstanding Young Scientist Research Fund under grant No. of NSF \# 19625514.}    \\  
\\
Institute of Theoretical Physics, 
Chinese Academy of Sciences \\ Beijing 100080,  China \\
e-mail: {\tt ylwu@itp.ac.cn } }

\maketitle

\begin{abstract}
 After presenting an optimistic prediction for the CP-violating parameter $\epsilon'/\epsilon$ 
in the standard model, we provide an analysis on new physics effects in $\epsilon'/\epsilon$.
\end{abstract}

\maketitle

\newpage

\section{Optimistic Prediction for $\epsilon'/\epsilon$ }

  The KTeV collaboration at Fermilab and NA48 collaboration at CERN have reported their more 
precise measurements on CP-violating ratio $\epsilon'/\epsilon$ in $K\rightarrow \pi \pi$ decays  
\begin{eqnarray}
Re(\epsilon'/\epsilon) & = & (28.0\pm 4.1) \cdot 10^{-4} \qquad KTeV\cite{KTEV}, \nonumber \\
Re(\epsilon'/\epsilon) & = & (18.5\pm 7.3) \cdot 10^{-3} \qquad NA48\cite{NA48}. \nonumber 
\end{eqnarray}

 The direct CP violation arises from the phase difference between 
the $K\rightarrow \pi\pi$ isospin amplitudes $A_{2}$ and $A_{0}$ and is defined as
\begin{equation}
Re(\epsilon'/\epsilon)\simeq \frac{\epsilon'}{\epsilon} = \frac{1}{\sqrt{2}|\epsilon|} 
Im \left(\frac{A_{2}}{A_{0}}\right)  
\end{equation}
Thus the ratio $\epsilon'/\epsilon$ can be simply rewritten as 
\begin{eqnarray}
 & & Re\left(\frac{\epsilon'}{\epsilon}\right)_{SM} \simeq 2.0 \cdot 10^{-3} 
\left(\frac{Im\lambda_{t}}{1.3\cdot 10^{-4}}\right) \\
& & 
\cdot \left( \frac{175 MeV}{m_{s}(1GeV)} \right)^{2} \left( \frac{-y_{6}B_{6}^{(1/2)}}{0.30}\right) 
\left( \frac{y_{B}}{1.5} \right)    \nonumber 
\end{eqnarray}
where $y_{B}$ is defined as $   
y_{B} \equiv 2 - r_{y}r_{B} + r_{0} $
with $r_{0}= (0.2-\Omega_{\eta-\eta'}-\Omega')/0.8$. The parameters 
$r_{y}$ and $r_{B}$ are defined as the following ratios 
$r_{y} \equiv -2y_{8}/(\alpha_{em}3y_{6})$ and  
$r_{B} = B_{8}^{(3/2)}/B_{6}^{(1/2)} $. Here the parameters $y_{6}$, $y_{8}$, 
$B_{6}^{(1/2)}$ and $B_{8}^{(3/2)}$ take values at the same energy scale ($\mu =1$ GeV). For 
conservative, we take 
\begin{equation}
 0.8\times 10^{-4} \leq Im\lambda_{t}  \leq 1.8\times 10^{-4} 
\end{equation}  
It is easy to check that the ratio $r_{y}$ only has a weak dependence on 
the scales $\mu$ and $\Lambda_{QCD}$ \cite{QCD2}  $r_{y} = 0.91\pm 0.09$ (at\  Leading\ Order),
$r_{y} = 0.98\pm 0.08$ (in\ NDR\ scheme) and $r_{y} = 1.28\pm 0.08$ (in\ HV\ scheme) 
when the energy scale $\mu$ ranges from 1 GeV to 2 GeV and the QCD scale $\Lambda_{QCD}$ 
from about 200 MeV to 400 MeV with $m_{t}(m_{t}) = 170$GeV.  
A large value for $B_{6}^{(1/2)}$ was emphasized in \cite{DM0,YLW}. A
chiral relation for $<Q_{6}>$  at $O(p^{2})$ order has been found to be useful\cite{YLW}
\begin{eqnarray}
 <Q_{6}> & = & (2m_{K}^{2}/m_{s}\Lambda_{\chi})^{2}f_{\pi}^{4} 
tr(\lambda_{6}\partial_{\mu}U\partial^{\mu}U^{\dagger})  \nonumber \\
& =  & 
 - (2m_{K}^{2}/m_{s}\Lambda_{\chi})^{2} <Q_{2}-Q_{1}>  \nonumber 
\end{eqnarray}
which implies that chiral loop corrections to the $O(p^{2})$ order terms will lead $<Q_{6}>$ to have 
the same enhancement as $<Q_{2}-Q_{1}>$. Therefore, when the other corrections, such as the $O(p^{4})$ 
terms at tree level and the nonfactorizable contributions from the terms $O(p^{0}/N_{c})$, are neglected, 
one has 
\[  B_{6}^{(1/2)} \sim B_{12}^{(1/2)} \equiv <Q_{2}-Q_{1}>/<Q_{2}-Q_{1}>^{tree} \] 
Recently, it has been shown \cite{DM2} that the chiral relation between $<Q_{6}>$ and 
$<Q_{2}-Q_{1}>$ is modified due to possible nonfactorizable chiral loop effects of 
the terms $O(p^{0}/N_{c})\sim Tr(\lambda_{6}U U^{\dagger})$ which vanishes at tree level. 
 At the chiral one-loop level, it has been found 
that $B_{12}^{(1/2)}(0.8GeV) = 3.5 $ \cite{DM1}.
Numerically, by including about $30\%$ reduction due to the nonfactorizable chiral loop effects of the 
terms $O(p^{0}/N_{c})$ and about $30\%$ corrections arising from possible higher order terms, 
we obtain the following range for $B_{6}^{(1/2)}$ at $\mu \simeq 0.8$ GeV
\begin{equation}
B_{6}^{(1/2)} = (1 - \Delta_{B})B_{12}^{(1/2)} \simeq 2.4 \pm 0.7  
\end{equation}
Note that the parameter $B_{8}^{(3/2)}$, unlike the parameter $B_{6}^{(1/2)}$, 
does not deviate significantly from the VSA value. Taking $r_{B} \simeq 0.5$, 
we arrive at the following constraint on $y_{B}$ 
\begin{eqnarray}
& & y_{B} = 1.55 \pm 0.14 \qquad (at\  Leading\ Order); \nonumber  \\ 
& &  y_{B} = 1.51 \pm 0.13\qquad (in\ NDR\ scheme); \nonumber  \\  
& & y_{B} =  1.36 \pm 0.13 \qquad (in\ HV\ scheme)  \nonumber 
\end{eqnarray}
For the coefficient function $y_{6}$, there remains a big uncertainty due to scheme- and 
$\Lambda_{QCD}$- dependence 
\begin{equation} 
 0.09\leq y_{6}(0.8GeV) \leq 0.24 
\end{equation}

With the above analyses, we arrive at the following prediction for $Re (\epsilon'/\epsilon)$ in the SM
\begin{equation}
 Re (\epsilon'/\epsilon) = 2.0 ^{+2.8}_{-1.3}\times 
10^{-3} \left( \frac{175 MeV}{m_{s}(1GeV)}\right)^{2} 
\end{equation}
Here the central value is consistent with the current world averaged value  
\begin{equation}
 Re (\epsilon'/\epsilon) = (2.1\pm 0.46)\times 10^{-3}  
\end{equation}
      
   Several groups\cite{DM0,DCP} have investigated the contributions to 
$\epsilon'/\epsilon$. With the current status of the SM predictions 
for the ratio $\epsilon'/\epsilon$, it may be too 
early to call for new physics due to large uncertainties of the relevant parameters. 
The SM with three families accommodates CP violation in quark decays in an elegant and economical 
manner, there are little evidence, however, that this explanation is unique. 
In fact, there are hints from cosmological arguments that it is not. In present theories with 
baryon-number-violating processes, the universe at very early times contains equal (thermal) 
populations of particles and antiparticles. How and when the currently observed excess of 
matter over antimatter is one of the key puzzles of cosmology. Attempts to explain this with 
only standard KM CP violation phase have so far not been successful. At this time we do not yet know the 
origin of the CP violations. In such a situation it is useful to make consistent hypotheses which 
can be either disproved by experiments or guide us to the origin of the effect.

\section{$\epsilon'/\epsilon$ in Two Higgs Doublet Model}

 One of the most popular origins of the effect is spontaneous breaking of the symmetry as was first 
pointed out by T.D. Lee. A complete model and systematic analysis has been 
worked out in ref.\cite{WW} by considering one of the simplest extensions of 
the standard model with an extra Higgs doublet (S2HDM). The model is based on two basic assumptions: 
(i) abandoning the {\it ad hoc} discrete symmetry and considering 
approximate flavor symmetry; (ii) CP violation originates solely 
from a single relative phase of the two vacuum expectation values (VEVs).
As a consequence, there exists a variety of new sources of CP violation. 
 
  The original Yukawa interaction has the form
\[ L_{Y} = \bar{\psi}_{L} (\Gamma_{1}^{D} \phi_{1} + \Gamma_{2}^{D}\phi_{2}) D_{R} 
+ \bar{\psi}_{L} (\Gamma_{1}^{U} \bar{\phi}_{1} + \Gamma_{2}^{U}\bar{\phi}_{2}) U_{R} \]
plus a similar term for leptons. 

Rewriting $L_{Y}$ in terms of 
the Higgs basis and the quark mass basis, 
we have $L_{Y} = (\sqrt{2}G_{F})^{1/2}(L_{1} + L_{2}) + H.c. $
with
\begin{eqnarray} 
& & L_{1} = \sqrt{2} ( H^{+} \sum_{i,j}^{3} 
 \xi_{d_{j}} m_{d_{j}} V_{ij} \bar{u}_{L}^{i} d^{j}_{R} \nonumber \\
& & - H^{-} \sum_{i,j}^{3}\xi_{u_{j}} m_{u_{j}} V^{\dagger}_{ij} 
\bar{d}_{L}^{i} u^{j}_{R} ) \nonumber \\ 
& & + H^{0} \sum_{i}^{3} (m_{u_{i}} \bar{u}_{L}^{i} u^{i}_{R} + 
m_{d_{i}} \bar{d}_{L}^{i} d^{i}_{R} ) \nonumber \\
& & +  \Phi \sum_{i}^{3} \xi_{d_{i}} m_{d_{i}} \bar{d}_{L}^{i} d^{i}_{R}  
+ \Phi^{\ast} \sum_{i}^{3} \xi_{u_{i}} m_{u_{i}} \bar{u}_{L}^{i} u^{i}_{R} 
 \nonumber
\end{eqnarray}
\begin{eqnarray} 
& & L_{2} =  \sqrt{2} ( H^{+} \sum_{i,j'\neq j}^{3} 
 V_{ij'} \mu^{d}_{j'j} \bar{u}_{L}^{i} d^{j}_{R} \nonumber \\
& & - H^{-}\sum_{i,j'\neq j}^{3} V^{\dagger}_{ij'}\mu^{u}_{j'j} 
\bar{d}_{L}^{i} u^{j}_{R} ) + \Phi \sum_{i\neq j}^{3} \mu^{d}_{ij}  
\bar{d}_{L}^{i} d^{j}_{R}  \nonumber \\
& & + \Phi^{\ast}\sum_{i\neq j}^{3} \mu^{u}_{ij} \bar{u}_{L}^{i} u^{j}_{R} 
 \nonumber
\end{eqnarray}
Where $H^{0}$ is the ``real" Higgs boson and the orthogonal state $\Phi\equiv (R + i I)$ 
forms a doublet with the charged Higgs $H^{\pm}$. The neutral mass eigenstates $H_{1}^{0}$, 
$H_{2}^{0}$, $H_{3}^{0}$ are related to $(R, \  H^{0}, \  I)$ by an orthogonal 
matrix $O^{H}$. 
 
The factors $\xi_{d_{j}}m_{d_{j}}$ ($\xi_{u_{j}}m_{u_{j}}$) arise primarily from diagonal elements
of $\Gamma_{1}^{D}$ ($\Gamma_{1}^{U}$) and $\Gamma_{2}^{D}$ ($\Gamma_{2}^{U}$) 
whereas the factors $\mu_{jj'}^{d}$ ($\mu_{jj'}^{u}$) arise from 
the small off-diagonal elements. In the physical mass eigenstate basis for all particles, 
there are four major sources of CP violation induced from the single CP violation phase $\delta$:

(1) \  The KM phase in the CKM matrix in $W^{\pm}$-boson and charged-Higgs
exchanges; 

(2) \  The new milliweak CP violation phases in the factors $\xi_{q}$ (q= u,d,c,s,t,b) 
in both charged and neutral Higgs exchange processes, which is independent of the KM phase; 

(3) \  The superweak effects due to phases in the factors $\mu_{ij}^{f}$. 
These yield CP violation in FCNE; 

(4)\  From the Higgs potential one derives the matrix $O^{H}$ that 
diagonalizes the $3\times 3$ neutral Higgs mass matrix. Even in the absence of fermions this 
$O^{H}$ may violate CP invariance.

  From the above sources of CP violation, the dominant contributions to 
the ratio $\epsilon'/\epsilon$ are 
\begin{eqnarray}
& & Re(\epsilon'/\epsilon) \equiv Re(\epsilon'/\epsilon)_{SM+H^{+}}  + 
Re(\epsilon'/\epsilon )_{H^{+}-Tree}  \nonumber \\
& & \qquad \   \  + \   Re(\epsilon'/\epsilon)_{H^{+}-LD} + 
Re (\epsilon'/\epsilon )_{H^{0}-Tree}  \nonumber \\ 
& &  \nonumber \\
& & \simeq Re\left(\frac{\epsilon'}{\epsilon}\right)_{SM} \left(\frac{\tilde{y}_{6}}{y_{6}}\right)
\left(\frac{\tilde{y}_{B}}{y_{B}}\right)
\nonumber \\
& &  + 2.3 \cdot 10^{-3} \left(\frac{Im (\xi_{s}^{\ast}\tilde{\xi}_{d})}{(11)^{2}}\right) 
\left(\frac{200GeV}{m_{H^{+}}} \right)^{2} \nonumber  \\
& &  \nonumber \\
& &  + (0.2-3.0)\times 10^{-3} Re\left(\epsilon_{H^{+}-LD}/\epsilon^{exp.}\right) \nonumber \\
& & \nonumber \\
& & - 0.88\cdot 10^{-4} \left( \frac{Im \tilde{\eta}^{(k)}_{d}}{10} \right) 
\left(\frac{125 GeV}{m_{H_{k}^{0}}}\right) \nonumber \\
& & \cdot \left(\sqrt{\frac{\Delta m_{K}^{H^{0}}}{\Delta m_{K}^{exp.}}}\right)
\left(\frac{Re X_{k,12}^{d\ast}}{\sqrt{Re(Y_{k,12}^{d\ast})^{2}}}\right) \nonumber 
\end{eqnarray}
with $\tilde{y}_{B} = 2 -\tilde{r}_{y} r_{B}$, $\tilde{r}_{y} = - 
2\tilde{y}_{8}/(3\tilde{y}_{6}\alpha_{em})$, $\tilde{\xi}_{d} = \xi_{d} + 
\xi_{u}^{\ast} m_{u}/m_{d}$, $\tilde{\eta}^{(k)}_{d} = 
\eta_{d}^{(k)} - \frac{m_{u}}{m_{d}}\eta_{u}^{(k)}$,  
$X_{k,12}^{d}  =  (\mu_{k,12}^{d} + \mu_{k,21}^{d \ast})/2$ and $ Y_{k,12}^{d}  = -i
(\mu_{k,12}^{d} - \mu_{k,21}^{d \ast})/2 $  
 
 In addition to the known features due to the induced KM source, 
one sees that the new sources through charged Higgs boson exchange can make a significant contribution 
to $\epsilon'/\epsilon$
\[
\frac{\epsilon'}{\epsilon}  \simeq  2\times 10^{-3}\ ,\   \   
\tan\beta \sim 8\  \left(\frac{m_{H^{+}}}{(200GeV)} \right) \]
without conflicting with other constraints. The indirect CP-violating paramter 
$\epsilon$ could be fitted by the new sources from 
box diagrams containing $H^{\pm}$. It may also receive significant contribution from superweak 
FCNE. The new sources could provide large CP violation in hyperon decays. 
CP asymmetry in the decay $b\rightarrow s\gamma $ may also be larger than in the 
standard model due to the new source. The resulting values for the electric dipole 
moments $D_{e}$ and $D_{n}$ of 
the electron and  neutron from the new sources could be close to the present limits.

\section{  $\epsilon'/\epsilon$ in Other Models}

In the  minimal supersymmetric standard model (MSSM), when abandoning flavor universality ansatze and 
considering approximate flavor symmetry
as well as assuming CP violating triliear Yukawa couplings, one has investigated the supersymmetric 
contributions to $\epsilon'/\epsilon$\cite{MSSMCP}. 
The relevant part of the effective hamiltonian is $H_{eff}^{\Delta S =1} = C_{8}Q_{8G}$  with
\begin{eqnarray}
C_{8} &=& \frac{\alpha_{s}\pi}{m_{\tilde{q}}^{2}} [ (\delta^{d}_{12})_{LL} (-\frac{1}{3}M_{3}(x)- 
3M_{4}(x) ) \nonumber \\
& & + (\delta^{d}_{12})_{LR}\frac{m_{\tilde{g}}}{m_{s}} ( -\frac{1}{3} - 3M_{2}(x) ) ]
\end{eqnarray}  
where $x= m_{\tilde{g}}^{2}/m_{\tilde{q}}^{2}$. Consequently,  for 
$|Im(\delta^{d}_{12})_{LR}^{2}|^{1/2} \simeq 10^{-5}$ and $m_{\tilde{q}} \simeq 500 GeV$, 
one has
\[ (\epsilon'/\epsilon)_{SUSY} \simeq 3\times 10^{-3} \]

 Recently, it was also shown\cite{XGH} that the contributions 
to the ratio $Re (\epsilon'/\epsilon)$ from anomalous gauge couplings could also be significant.


\begin{thebibliography}{99}
\bibitem{KTEV} H. Nguyen, in this proceedings;\\ A. Alavi-Harati {\it et al.}, Phys. Rev. Lett. {\bf 83} 
(1999) 22.
\bibitem{NA48} S. Palestini, in this proceedings; \\ V. Fanti {\it et al.}, hep-ex/9909022.
\bibitem{QCD2}  A. Buras, M. Jamin and M. Lautenbacher, Nucl. Phys. {\bf B408}, 209 (1993);
 M. Cuichini et al., Nucl. Phys. {\bf B415}, 403 (1994).
\bibitem{DM0} J. Heinrich, E.A. Paschos, J.-M. Schwarz, and Y.L. Wu, Phys. Lett. {\bf B279}, 140 (1992);
\\ E.A. Paschos, Proceedings of the 27th Lepton-Photon Symposium, Beijing, China 1995, edited 
by Z.P. Zheng and H.S. Chen, (World Scientific, Singapore, 1996).
\bibitem{YLW}Y.L. Wu, Intern. J. of Mod. Phys. A7 (1992) 2863;  Y.L. Wu
in: Proceedings of 26TH Int. Conf. on High Energy Physics, 
Dallas, USA, 1992, PP 506, edited by James R. Sanford.
\bibitem{DM2} T. Hambye, G.O. K\"{o}hler, E.A. Paschos, P.H. Soldan, and W.A. Bardeen, 
Phys. Rev. {\bf D58}, 014017 (1998).
\bibitem{DM1}T. Hambye, G.O. K\"{o}hler, and P.H. Soldan, hep-ph/9902334; J.P. Fatelo and J.-M. G\'{e}rard, 
Phys. Lett. {\bf B257}, 191 (1991).
\bibitem{DCP} M.Ciuchini, Nucl. Phys. B (Proc. Suppl. 59 (1997) 149; A.J. Buras {\it et al.}, 
Nucl. Phys. {\bf B370} (1993) 209; S. Bertolini {\it et al.}, Nucl. Phys. {\bf B514} (1998) 93; \\
S. Bosch {\it et al.}, hep-ph/9904408; \\ T. Hambye {\it et al.}, hep-ph/9906434; \\ A.A. Bel'kov {\it 
et al.}, hep-ph/9907335.
\bibitem{WW}  Y.L. Wu and L. Wolfenstein, Phys. Rev. Lett. {\bf 73}, 1762 (1994); 
L. Wolfenstein and Y.L. Wu, Phys. Rev. Lett., {\bf 73}, 2809 (1994); \\
 Y.L. Wu,  hep-ph/9404241, hep-ph/9903363  
\bibitem{MSSMCP} A. Masiero and H. Murayama, hep-ph/9903363; 
K.S. Babu, B. Dutta and R.N. Mohapatra, hep-ph/9905464; S.Khalil and T. Kobayashi, hep-ph/9906374; \\
S. Baek, J.-H. Jang, P. Ko and J.H. Park, hep-ph/9907572; 
R. Barbieri, R. Contino and A. Strumia, hep-ph/9908255; 
G. Eyal, A. Masiero, Y. Nir and L. Silvestrini, hep-ph/9908382.
\bibitem{XGH} X.G. He, hep-ph/9903242.

\end{thebibliography}
\end{document}